\documentstyle[prb,aps,psfig,twocolumn,
]{revtex}
\begin{document}

\onecolumn

\title{Analytical Hartree-Fock gradients with respect to  the cell parameter
for systems periodic in three dimensions}
\author{K. Doll} 
\address{Institut f\"ur Mathematische Physik, TU Braunschweig,
Mendelssohnstra{\ss}e 3, D-38106 Braunschweig, Germany}
\author{R. Dovesi, R. Orlando} 
\address{Dipartimento di Chimica IFM, Universit\`a
 di Torino, Via Giuria 5, I-10125 Torino, Italy}

\maketitle

\begin{abstract}
Analytical Hartree-Fock gradients with respect to the cell parameter have
been implemented in the electronic structure code CRYSTAL, for the
case of three-dimensional periodicity. The code is based on
Gaussian type orbitals, and the summation of the Coulomb energy is
performed with the Ewald method.
It is shown that a high accuracy of the cell gradient can be achieved.
\end{abstract}

\pacs{ }

\narrowtext
\section{Introduction}

Electronic structure codes are nowadays widely used by theoreticians and
also by a growing number of experimentalists. 
One of the targets is the calculation of the total energy and the
structural optimization of large systems.
It is well known that this is greatly facilitated by the availability
of analytical gradients, and thus the coding of gradients has
become an important part of code development. 

In the molecular case, the geometrical parameters
that must be optimized, are
the nuclear coordinates. This field was pioneered by Pulay\cite{Pulay};
however it should be mentioned that
the theory had already been derived earlier independently\cite{Bratoz}.
Meanwhile, numerous review articles about analytical derivatives have appeared
\cite{PulayAdv,PulayChapter,Helgaker,HelgakerJorgensen1992,SchlegelYarkony,PulayYarkony,Schlegel2000}. 

In the case of periodic systems, the cell
dimensions are an additional set of parameters. 
Nowadays, nearly all solid-state codes compute the total energy 
with density functional methods. However, because of the success
of hybrid functionals, which use an admixture of Fock exchange, 
there is a growing interest in codes which compute the Fock exchange.
CRYSTAL\cite{CRYSTALbuch} 
was originally a code for pure Hartree-Fock calculations.
In the past decade, density functional calculations for all types of
functionals have become possible as well with this code. 

The code is based on Gaussian type orbitals, and most of the
contributions to the total energy rely
on the Hartree-Fock formulation. Therefore, the Hartree-Fock gradients
with respect to the cell parameter were the first step to make
gradients with respect to the cell parameter available.
In this article, we will report on the implementation of gradients
at the Hartree-Fock level,
with respect to the cell parameter, for systems periodic in three
dimensions. 

The first gradients with respect to the cell parameter, 
at the Hartree-Fock level,
were for systems periodic in one dimension\cite{Teramae198384}. Meanwhile,
various groups have implemented these 
gradients in one dimension\cite{Jacquemin,Hirata1997} or
in two dimensions\cite{Tobita2003}. For the general case, a strategy
to compute cell parameter derivatives (and thus the stress tensor)
was suggested with point charges \cite{Kudin2000}, and
an algorithm for structural optimization, based on 
redundant internal coordinates was proposed\cite{Kudin2001}.

All these codes use a real space approach, where all the summations
are performed in direct space. As an acceleration tool, the fast
multipole method is applied in various cases. The cell parameter gradient
is then obtained essentially by multiplying the contributions to the
nuclear gradients with the appropriate factors. 

In contrast, the CRYSTAL code uses, in the case of three-dimensional 
periodicity,
the Ewald method which is a combination of direct and reciprocal lattice
summations\cite{Ewald}. 
This means, that besides some contributions which have to be
multiplied with trivial factors, there will also be additional derivatives,
because various parameters in the Ewald scheme, and the reciprocal lattice
vectors, depend on the cell parameters. The Hartree-Fock
gradients with respect to
nuclear coordinates were implemented earlier\cite{IJQC,CPCarticle}, 
and after the
implementation of a tool for structural optimization, 
it was demonstrated
that an efficient geometry optimization of large systems with any periodicity
can be performed\cite{MimmoArcoetal}.

In this article, we describe the formalism used in the CRYSTAL 
code for the cell parameter gradients, 
and present results from tests on various systems. The article
is structured as follows: first, the variables are defined. Then
the integrals and their derivatives with respect to the cell parameters
are discussed, and in the following section
the derivative of the total energy is given. Finally, some examples
are used as an illustration of the formalism.

\section{Variables}

The primitive cell is given by three vectors: 
$\vec a_1$, $\vec a_2$ and $\vec a_3$, and
the derivatives with respect to the cell parameters $a_{ij}$ have
been coded. $a_{ij}$ are defined in such a way that $a_{11}=a_{1x}$ 
is the
x-component of $\vec a_{1}$, $a_{12}=a_{1y}$ the $y$-component of $\vec a_1$, 
and $a_{13}=a_{1z}$ the $z$-component of $\vec a_1$, i.e.:

\begin{eqnarray}
\left(
\begin{array}{c} 
\vec a_1\\
\vec a_2\\
\vec a_3\\
\end{array}
\right)=
\left(
\begin{array}{c} 
a_{1x} \ a_{1y} \ a_{1z}\\
a_{2x} \ a_{2y} \ a_{2z}\\
a_{3x} \ a_{3y} \ a_{3z}\\
\end{array}
\right)=
\left(
\begin{array}{c} 
a_{11} \ a_{12} \ a_{13}\\
a_{21} \ a_{22} \ a_{23}\\
a_{31} \ a_{32} \ a_{33}\\
\end{array}
\right)
\end{eqnarray}

A point $\vec g$ of the direct lattice is defined 
as $\vec g=n_1 \vec a_1+n_2 \vec a_2+n_3 \vec a_3$, with $n_1, n_2, n_3$
being integer numbers. 
We assume to have $N$ atoms in the unit cell.
The position 
of an atom $c$ in a cell at the origin 
(i.e. $\vec g=\vec 0$) is given as
$\vec A_{c}=f_{c,1}\vec a_1+f_{c,2}\vec a_2+f_{c,3}\vec a_3$, and then
in cell $\vec g$ the position will be:

$\vec A_{c}+\vec g=(f_{c,1}+n_{\vec g,1})\vec a_1+(f_{c,2}+
n_{\vec g,2})\vec a_2+(f_{c,3}+n_{\vec g,3})\vec a_3$

We have used an additional index, i.e. 
$n_{\vec g,1}$ means factor $n_1$ of the lattice vector $\vec g$.
The cartesian $t$ component (with $t$ being $x$, $y$ or $z$)
of the vector $\vec A_{c}+\vec g$, indicated
as $A_{c,t}+g_{t}$, is thus

$A_{c,t}+g_{t}=\sum_{m=1}^3 (f_{c,m}+n_{\vec g,m})a_{mt}$

As all the integrals depend on the position of the nuclei, the derivatives
of the nuclear coordinates with respect to the cell parameters are
required:

\begin{eqnarray}
\frac{\partial A_{c,t}+g_{t}}{\partial a_{ij}}=
\sum_{m=1}^3 (f_{c,m}+n_{\vec g,m})\delta_{im}\delta_{jt}=
(f_{c,i}+n_{\vec g,i})\delta_{jt}
\end{eqnarray}

with the Kronecker symbol $\delta_{jt}$.

In the following, we will use the notation $\vec A_{c}$ to indicate the
position of
atom $c$.
However, we also need to define basis
functions $\phi_{\mu}$ centered at a certain nucleus, where $\mu$ runs
over all the basis functions. For example, basis functions $\mu=1,...,5$
might be centred at atom 1, basis functions $\mu=6,...,15$ 
at atom 2 and so on.
It is thus trivial to assign a certain
atom number $c$ to the basis function $\mu$: $c=c(\mu)$. We could thus use
the notation $\vec A_{c(\mu)}$, but will instead use the simplified
notation $\vec A_{\mu}$.
To avoid confusion, Greek indices are used in this case, i.e.
$\vec A_{\mu}=\vec A_{c(\mu)}=\vec A_{c}$.

The basis functions used are real spherical Gaussian type functions,
and we will use the notation $\phi_{\mu}(\vec r-\vec A_{\mu}-\vec g)$.

The crystalline orbitals are linear combinations of Bloch functions

\begin{equation}
\Psi_n(\vec r, \vec k)=\sum_{\mu} a_{\mu n}(\vec k)\psi_{\mu}(\vec r, \vec k)
\end{equation}

which are expanded in terms of real spherical Gaussian type functions

\begin{equation}
\psi_{\mu}(\vec r, \vec k)=\sum_{\vec g}
\phi_{\mu}(\vec r-\vec A_{\mu}-\vec g)
{\rm e}^{{\rm i} \vec k\vec g}
\end{equation}

The spin-free density matrix in reciprocal space is defined as

\begin{eqnarray}
P_{\mu\nu}(\vec k)=2\sum_{n}a_{\mu n}(\vec k)a_{\nu n}^*(\vec k)
\Theta(\epsilon_F-\epsilon_n(\vec k))
\end{eqnarray}

with the Fermi energy $\epsilon_F$ and the Heaviside function $\Theta$.
In the case of unrestricted Hartree-Fock (UHF) \cite{Apra}, 
we use the notation

\begin{equation}
\Psi_n^{\uparrow}(\vec r, \vec k)=\sum_{\mu} a^{\uparrow}_{\mu n}(\vec k)
\psi_{\mu}(\vec r, \vec k)
\end{equation}
\begin{equation}
\Psi_n^{\downarrow}(\vec r, \vec k)=\sum_{\mu} a^{\downarrow}_{\mu n}(\vec k)
\psi_{\mu}(\vec r, \vec k)
\end{equation}

for the crystalline orbitals with up and down spin, respectively.
We define
the density matrices for spin up and spin down as follows:

\begin{eqnarray}
P_{\mu\nu}^{\uparrow}(\vec k)=\sum_{n}a_{\mu n}^{\uparrow}(\vec k)a_{\nu n}
^{* \uparrow}(\vec k)
\Theta(\epsilon_F-\epsilon_n^{\uparrow}(\vec k))
\end{eqnarray}

and 

\begin{eqnarray}
P_{\mu\nu}^{\downarrow}(\vec k)=\sum_{n}a_{\mu n}^{\downarrow}(\vec k)
a_{\nu n}
^{* \downarrow}(\vec k)
\Theta(\epsilon_F-\epsilon_n^{\downarrow}(\vec k))
\end{eqnarray}

In the following, $P_{\mu\nu}$ refers to the sum
$P_{\mu\nu}^{\uparrow}+P_{\mu\nu}^{\downarrow}$ in the UHF case.
The density matrices in real space 
$P_{\mu\vec 0\nu\vec g}, P_{\mu\vec 0\nu\vec g}^{\uparrow}, 
P_{\mu\vec 0\nu\vec g}^{\downarrow}$
are obtained by Fourier transformation of the corresponding reciprocal
space quantity.

\section{Integrals and their derivatives}

\label{Integralsection}


\subsection{Nuclear-nuclear repulsion energy}
The Ewald energy of the nuclear repulsion $E^{\rm NN}$ is obtained as

\begin{eqnarray} & 
E^{\rm NN}= &
\frac{1}{2}
\sum_{a,b} Z_a Z_b \Phi(\vec A_{b}-\vec A_{a})= \nonumber \\
 & &
\frac{1}{2}
\sum_{a,b} Z_a Z_b \left( -\frac{\pi}{\gamma V}+\sum_{\vec h}^{'} 
\frac{{1-\rm erf}
(\sqrt\gamma|\vec A_{b}-\vec A_{a}- \vec h|)}{|\vec A_{b}-
\vec A_{a}- \vec h|}
+\frac{4\pi}{V}\sum_{\vec K}^{'}\frac{1}{\vec K^2}\exp\left(
-\frac{\vec K^2}{4\gamma}+{\rm i} \vec K(\vec A_{b}-\vec A_{a})
\right) \right)
\end{eqnarray}

with the nuclear charge $Z_a$, and
$\Phi(\vec r-\vec A_{a})$ being the Ewald potential function $\Phi$, 
as defined in 
reference \onlinecite{VicCoulomb}:

\begin{eqnarray} &  
\Phi(\vec r-\vec A_{a}) & = -\frac{\pi}{\gamma V}+\sum_{\vec h}^{'} 
\frac{{1-\rm erf}
(\sqrt\gamma|\vec r-\vec A_{a}-\vec h|)}{|\vec r-\vec A_{a}- \vec h|}
+\frac{4\pi}{V}\sum_{\vec K}^{'}\frac{1}{\vec K^2}\exp\left(
-\frac{\vec K^2}{4\gamma}+{\rm i} \vec K(\vec r-\vec A_{a})\right)
\nonumber \\ & &
= -\frac{\pi}{\gamma V}+\sum_{\vec h}^{'} \Phi_{\vec h}(\vec r-\vec A_{a})
+\sum_{\vec K}^{'}
\Phi_{\vec K}(\vec r-\vec A_{a})
\end{eqnarray}

$\vec h$ are the direct lattice vectors, $\vec K$ the reciprocal
lattice vectors. $V$ is the volume, $\gamma$ is a screening parameter 
which was optimized\cite{VicCoulomb} to be $\gamma=(2.8/V^{1/3})^2$, in
the three dimensional case.
The prime in the direct lattice summation 
indicates that the summation includes all values of the
direct lattice vector $\vec h$, with the exception
of the case when $|\vec r-\vec A_{a}-\vec h|$ vanishes. 
In this case, the term $\frac{1}{|\vec r-\vec A_{a}-\vec h|}$ is omitted
from the sum. In the 
reciprocal lattice series, the prime indicates that
all terms with $\vec K \neq \vec 0$
are included.

The error function is defined as

\begin{eqnarray}
{\rm erf}(p)=\frac{2}{\sqrt{\pi}}\int_0^p \exp (-u^2)du
\end{eqnarray}

We consider the Ewald potential as being dependent on the variables
$\vec A_{c}$ , $\vec h$, 
$V$, $\gamma$ and
$\vec K$.
The derivative with respect to the cell parameters thus requires
derivatives with respect to the centers $\vec A_{c}$ and the lattice vectors
$\vec h$
which are similar to the nuclear gradients and
have to be multiplied with a trivial factor.
In addition, the
Ewald potential depends on the cell parameters $a_{ij}$ through
the volume $V$, the
screening parameter $\gamma$, and the reciprocal lattice vectors $\vec K$. 
We obtain:

\begin{eqnarray} & &
\frac{\partial E^{\rm NN}}{\partial a_{ij}}= 
\sum_{c,t}\frac{\partial E^{\rm NN}}{\partial A_{c,t}}\frac{\partial A_{c,t}}
{\partial a_{ij}}+
\sum_t\frac{\partial E^{\rm NN}}{\partial h_t}\frac{\partial h_t}
{\partial a_{ij}}+ 
\nonumber \\ & &
\frac{\partial E^{\rm NN}}{\partial V}\frac{\partial V}
{\partial a_{ij}}+
\frac{\partial E^{\rm NN}}{\partial \gamma}
\frac{\partial \gamma}{\partial a_{ij}}+
\sum_t \frac{\partial E^{\rm NN}}{\partial K_t}
\frac{\partial K_t}{\partial a_{ij}}
=\sum_c\frac{\partial E^{\rm NN}}{\partial A_{c,j}}f_{c,i}+
\sum_{\vec h \ne \vec 0}^{'}
\frac{\partial E^{\rm NN}}{\partial h_j}n_{\vec h,i}+ 
\nonumber \\ & &
\frac{\partial E^{\rm NN}}{\partial V}\frac{\partial V}
{\partial a_{ij}}+
\frac{\partial E^{\rm NN}}{\partial \gamma}
\frac{\partial \gamma}{\partial a_{ij}}+
\sum_t \frac{\partial E^{\rm NN}}{\partial K_t}
\frac{\partial K_t}{\partial a_{ij}}
\end{eqnarray}

$K_t$, with $t=1,2,3$,
are the components $K_1=K_x, K_2=K_y, K_3=K_z$ of $\vec K$.
The derivatives of the parameters 
$V$, $\gamma$, $\vec K$ with respect to $a_{ij}$ are straightforward:

\subsubsection{Derivative of the volume}

The volume is obtained as the cross product of the cell parameters:

$V=\vec a_1(\vec a_2\times \vec a_3)$

Thus, the derivatives $\frac{\partial V}{\partial a_{ij}}$, 
are, for example, obtained as:

\begin{eqnarray}
\frac{\partial V}{\partial a_{1x}}=a_{2y}a_{3z}-a_{2z}a_{3y}\\
\frac{\partial V}{\partial a_{1y}}=a_{2z}a_{3x}-a_{2x}a_{3z}\\
\frac{\partial V}{\partial a_{2x}}=a_{3y}a_{1z}-a_{3z}a_{1y}
\end{eqnarray}

The remaining components can easily be obtained by cyclic permutation:
$1\rightarrow 2, 2 \rightarrow 3, 3\rightarrow 1$, or
$x\rightarrow y, y \rightarrow z, z\rightarrow x$.

\subsubsection{Derivative of the screening parameter}

The derivative is straightforward:

\begin{eqnarray}
\frac{\partial \gamma}{\partial a_{ij}}=
\frac{\partial \gamma}{\partial V}\frac{\partial V} {\partial a_{ij}}=
-\frac{2\gamma}{3V}\frac{\partial V}{\partial a_{ij}}
\end{eqnarray}

\subsubsection{Derivative of the reciprocal lattice vectors}

The reciprocal lattice vectors $\vec K$ can be expressed as

\begin{eqnarray}
\vec K=n_1 \vec b_1 + n_2 \vec b_2 + n_3 \vec b_3
\end{eqnarray}

with the primitive vectors $\vec b_i$ of the reciprocal lattice
defined as:

\begin{eqnarray}
\vec b_1=\frac{2\pi}{V}{\vec a_2 \times \vec a_3} \ ; \ \
\vec b_2=\frac{2\pi}{V}{\vec a_3 \times \vec a_1} \ ; \ \
\vec b_3=\frac{2\pi}{V}{\vec a_1 \times \vec a_2}
\end{eqnarray}

We thus need to evaluate the derivatives of the vectors $\vec b_i$ with 
respect to the cell parameters. A few examples are given below:

\begin{eqnarray}
\frac{\partial \vec b_1}{\partial a_{1x}}=
\frac{\partial}{ \partial a_{1x}}
\left(  \frac{2\pi}{V} \left(
\begin{array}{c} 
 a_{2y}a_{3z}-a_{3y}a_{2z}  \\
 a_{2z}a_{3x}-a_{3z}a_{2x} \\
 a_{2x}a_{3y}-a_{3x}a_{2y} \\
\end{array}
\right)\right)
\end{eqnarray}

and thus 

\begin{eqnarray}
\frac{\partial \vec b_1}{\partial a_{1x}}=
-\frac{\partial V}{\partial a_{1x}}\frac{\vec b_1}{V}
\end{eqnarray}

or

\begin{eqnarray}
\frac{\partial \vec b_1}{\partial a_{2x}}=
-\frac{\partial V}{\partial a_{2x}}\frac{\vec b_1}{V}
+\frac{2\pi}{V}
\left(
\begin{array}{c} 
0 \\ -a_{3z} \\ a_{3y} 
\end{array}
\right)
\end{eqnarray}

or

\begin{eqnarray}
\frac{\partial \vec b_2}{\partial a_{1x}}=
-\frac{\partial V}{\partial a_{1x}}\frac{\vec b_2}{V}
+\frac{2\pi}{V}
\left(
\begin{array}{c} 
0 \\ a_{3z} \\ -a_{3y}
\end{array}
\right)
\end{eqnarray}

Again, all the other derivatives can be obtained by cyclic permutation.
The individual derivatives are thus:

\begin{eqnarray}
 & &
\frac{\partial E^{\rm NN}}{\partial V}=
\frac{1}{2}
\sum_{a,b} Z_a Z_b \left( \frac{\pi}{\gamma V^2}
-\frac{4\pi}{V^2}\sum_{\vec K}^{'}\frac{1}{\vec K^2}\exp\left(
-\frac{\vec K^2}{4\gamma}+{\rm i} \vec K(\vec A_{b}-\vec A_{a})
\right) \right)
\end{eqnarray}

\begin{eqnarray}
 & &
\frac{\partial E^{\rm NN}}{\partial \gamma}=
\frac{1}{2}
\sum_{a,b} Z_a Z_b \left( \frac{\pi}{\gamma^2 V}
-\sum_{\vec h}^{'}\frac{1}{\sqrt {\pi\gamma}}
{\exp(-\gamma(\vec A_{b}-\vec A_{a}- \vec h)^2)}
+\frac{\pi}{\gamma^2 V}\sum_{\vec K}^{'}\exp\left(
-\frac{\vec K^2}{4\gamma}+{\rm i} \vec K(\vec A_{b}-\vec A_{a})
\right) \right)
\end{eqnarray}

\begin{eqnarray}
 & &
\frac{\partial E^{\rm NN}}{\partial K_t}=
\frac{1}{2}
\sum_{a,b} Z_a Z_b 
\frac{4\pi}{V}\sum_{\vec K}^{'}\left(\frac{-2K_t}{(\vec K^2)^2}
+\frac{1}{\vec K^2}\left(-\frac{K_t}{2\gamma}+
{\rm i}(A_{b,t}-A_{a,t})\right)\right)\exp\left(
-\frac{\vec K^2}{4\gamma}+{\rm i} \vec K(\vec A_{b}-\vec A_{a})
\right)
\end{eqnarray}

The partial derivative with respect to the nuclear positions is just
the nuclear gradient which is already available in the code:

\begin{eqnarray} 
\frac{\partial E^{\rm NN}}{\partial A_{c,t}} & = &
\frac{1}{2}
\sum_{a} Z_a Z_c 
\left(2  \sum_{\vec h}^{'} (A_{c,t}-A_{a,t}-h_t)\left(
-\frac{1-{\rm erf}(\sqrt\gamma|\vec A_{c}-\vec A_{a}- \vec h|)}
{|\vec A_{c}-\vec A_{a}- \vec h|^3}
-\frac{2\sqrt \gamma}{\sqrt \pi}
\frac{\exp(-\gamma(\vec A_{c}-\vec A_{a}- \vec h)^2)}
{|\vec A_{c}-\vec A_{a}- \vec h|^2}\right)
\right. \nonumber \\ & & \left.
+\frac{4\pi}{V}\sum_{\vec K}^{'}
\frac{{\rm i} K_t}{\vec K^2}
\exp\left(
-\frac{\vec K^2}{4\gamma}\right)
\left(\exp\left({\rm i} \vec K(\vec A_{c}-\vec A_{a})\right)
     -\exp\left({\rm -i} \vec K(\vec A_{c}-\vec A_{a})\right)
\right)
\right) 
\end{eqnarray}

The derivatives with respect to the direct lattice vectors $\vec h$
are obtained as:

\begin{eqnarray} 
\frac{\partial E^{\rm NN}}{\partial h_t} & = &
\frac{1}{2}\sum_{a,b} Z_a Z_b 
\sum_{\vec h\ne \vec 0} (A_{b,t}-A_{a,t} -h_t)\left(
\frac{1-{\rm erf}(\sqrt\gamma|\vec A_{b}-\vec A_{a}-\vec h|)}
{|\vec A_{b }-\vec A_{a}- \vec h|^3}
+\frac{2\sqrt \gamma}{\sqrt \pi}
\frac{\exp(-\gamma(\vec A_{b}-\vec A_{a}- \vec h)^2)}
{|\vec A_{b}-\vec A_{a}- \vec h|^2}\right)
\end{eqnarray}

\subsection{Overlap integral}

The fundamental integral is the overlap integral. It is defined as

\begin{eqnarray} & &
S_{\mu\vec {0}\nu\vec {g}}= 
\int \phi_{\mu}
(\vec r - \vec A_{\mu})
\phi_{\nu}(\vec r - \vec A_{\nu}-\vec g){\rm d^3r}
\end{eqnarray}

Its derivative with respect to the cell parameters $a_{ij}$ is thus

\begin{eqnarray} & &
\frac{\partial S_{\mu\vec {0}\nu\vec {g}}}{\partial a_{ij}}=
\sum_{t=1}^3
\left(
\frac{\partial S_{\mu\vec {0}\nu\vec {g}}}{\partial A_{\mu,t}}
\frac{\partial A_{\mu,t}}{\partial a_{ij}}+
\frac{\partial S_{\mu\vec {0}\nu\vec {g}}}{\partial (A_{\nu,t}+g_{t})}
\frac{\partial (A_{\nu,t}+g_{t})}{\partial a_{ij}}\right)
\end{eqnarray}

where $t$ represents the summation over $x,y,z$.

Exploiting translational invariance, we obtain

\begin{eqnarray}
\label{overlapequation}
 & &
\frac{\partial S_{\mu\vec {0}\nu\vec {g}}}{\partial a_{ij}}=
\sum_{t=1}^3
\frac{\partial S_{\mu\vec {0}\nu\vec {g}}}{\partial A_{\mu,t}}
\left( \frac{\partial A_{\mu,t}}{\partial a_{ij}}-
\frac{\partial (A_{\nu,t}+g_{t})}
{\partial a_{ij}}\right)= \nonumber \\ & & 
\sum_{t=1}^3
\frac{\partial S_{\mu\vec {0}\nu\vec {g}}}{\partial A_{\mu,t}}
\left( f_{\mu,i}-f_{\nu,i}-n_{\vec g,i} \right)
\delta_{jt}= \nonumber \\ & & 
\frac{\partial S_{\mu\vec {0}\nu\vec {g}}}{\partial A_{\mu,j}}
\left( f_{\mu,i}-f_{\nu,i}-n_{\vec g,i} \right)
\end{eqnarray}

The derivative 
$\frac{\partial S_{\mu\vec {0}\nu\vec {g}}}{\partial A_{\mu ,j}}$
is identical to the one which is required for the calculation of the
gradient with respect to nuclear positions
and thus it only needs to be multiplied with a
trivial factor to obtain the derivative with respect to the cell
parameters.

\subsection{Kinetic energy integrals}

The evaluation of the kinetic energy integrals leads to 
a combination of overlap integrals:

\begin{eqnarray} 
& & 
T_{\mu\vec 0\nu\vec g}=
\int \phi_{\mu}(\vec r-\vec A_{\mu})
\left(-\frac{1}{2}\Delta_{\vec r}\right)
\phi_{\nu}(\vec r-\vec A_{\nu}-\vec g)
{\rm d^3r}
\end{eqnarray}

When computing the derivatives with respect to the cell parameters,
we thus obtain in complete analogy to equation \ref{overlapequation}:

\begin{eqnarray} 
\frac{\partial T_{\mu\vec {0}\nu\vec {g}}}{\partial a_{ij}}=
\frac{\partial T_{\mu\vec {0}\nu\vec {g}}}{\partial A_{\mu,j}}
\left( f_{\mu,i}-f_{\nu,i}-n_{{\vec g},i} \right)
\end{eqnarray}

\subsection{Nuclear attraction integrals}

The nuclear attraction integrals are defined as

\begin{eqnarray} & & 
N_{\mu\vec 0\nu\vec g}= -\sum_a Z_a\int 
\phi_{\mu}(\vec r-\vec A_{\mu})\Phi(\vec r-\vec A_{a})
\phi_{\nu}(\vec r-\vec A_{\nu}-\vec g) {\rm d^3r}
\end{eqnarray}

It has been explicitly evaluated in reference \onlinecite{VicCoulomb}.
The derivative with respect to the cell parameters consists now
of the following contributions: there are derivatives with respect
to three centers $\vec A_{\mu}$, $\vec A_{\nu}+\vec g$ and 
$\vec A_{a}$. 
Similar to the nuclear-nuclear repulsion,
derivatives with
respect to $V$, $\gamma$ and $\vec K$ are required.
These derivatives are similar to the 
ones appearing when evaluating the nuclear-nuclear gradient. 

The derivatives with respect to the direct lattice vectors can again
be obtained by using the nuclear gradients, and the rule:

\begin{eqnarray}
\sum_{t=1}^3
\frac{\partial N_{\mu\vec 0\nu\vec g}}{\partial A_{\mu,t}}
\frac{\partial A_{\mu,t}}{a_{ij}}
=\frac{\partial N_{\mu\vec 0\nu\vec g}}{\partial A_{\mu,j}}f_{\mu,i}
\end{eqnarray}

The part of the derivative coming from the center 
$(\vec A_{\nu}+\vec g)$ is
obtained in the same way, and the derivative with respect to 
$\vec A_{a}$ in $\Phi(\vec r-\vec A_{a})$
is obtained from translational invariance. All these nuclear gradients
simply need to be multiplied with the proper factors, for 
$\vec A_{\mu}$,
$\vec A_{\nu}+\vec g$ and $\vec A_{a}$.

We can thus view the nuclear attraction integrals $N_{\mu\vec 0\nu\vec g}$
as a function of the variables $\vec A_{\mu}$, $\vec A_{\nu}+ \vec g$,
$\vec A_{a}$, $\vec h$, $V$, $\gamma$ and $\vec K$.
As a whole, we obtain therefore for the derivative:

\begin{eqnarray} & &
\frac{\partial N_{\mu\vec 0\nu\vec g}}{\partial a_{ij}}= 
\frac{\partial N_{\mu\vec 0\nu\vec g}}{\partial A_{\mu,j}}f_{\mu,i}+
\frac{\partial N_{\mu\vec 0\nu\vec g}}{\partial A_{\nu,j}}
(f_{\nu,i}+n_{\vec g,i})
-\left(\frac{\partial N_{\mu\vec 0\nu\vec g}}{\partial A_{\mu,j}}+
\frac{\partial N_{\mu\vec 0\nu\vec g}}{\partial A_{\nu,j}}\right)
f_{a,i} \nonumber \\ & &
-Z_a \sum_{\vec h\ne \vec 0}^{'}\int
\phi_{\mu}(\vec r-\vec A_{\mu})\frac{\partial \Phi_{\vec h}
(\vec r-\vec A_{a})}{\partial {h_j}}n_{\vec h,i}
\phi_{\nu}(\vec r-\vec A_{\nu}-\vec g) {\rm d^3r}
+\frac{\partial N_{\mu\vec 0\nu\vec g}}{\partial V}\frac{\partial V}
{\partial a_{ij}}+
\frac{\partial N_{\mu\vec 0\nu\vec g}}{\partial \gamma}
\frac{\partial \gamma}{\partial a_{ij}}+
\sum_t \frac{\partial N_{\mu\vec 0\nu\vec g}}{\partial K_t}
\frac{\partial K_t}{\partial a_{ij}}
\end{eqnarray}

Translational invariance can be exploited for
the following contribution:

\begin{eqnarray} & & 
\int
\phi_{\mu}(\vec r-\vec A_{\mu})\frac{\partial \Phi_{\vec h}
(\vec r-\vec A_{a})}{\partial {h_j}}n_{\vec h,i}
\phi_{\nu}(\vec r-\vec A_{\nu}-\vec g) {\rm d^3r}= \nonumber \\  & & 
-\int \frac{\partial \phi_{\mu}(\vec r-\vec A_{\mu})}{\partial A_{\mu,j}}
\Phi_{\vec h}(\vec r-\vec A_{a})
\phi_{\nu}(\vec r-\vec A_{\nu}-\vec g)n_{\vec h,i} {\rm d^3r}
-\int \phi_{\mu}(\vec r-\vec A_{\mu})\Phi_{\vec h}(\vec r-\vec A_{a})
\frac{\partial\phi_{\nu}(\vec r-\vec A_{\nu}-\vec g)}{\partial A_{\nu,j}}
n_{\vec h,i}{\rm d^3r}
\end{eqnarray}

\subsection{Multipolar integrals}

In the expression of the total energy, a term with a factor

\begin{eqnarray} 
\eta_l^m(\rho_c;\vec A_{c})=\int \rho_c(\vec r) 
X_l^m(\vec r-\vec A_{c}){\rm d^3r} 
\end{eqnarray}

appears,
with $X_l^m$ being regular solid harmonics \cite{VicCoulomb}.
The charge $\rho_c(\vec r)$ is defined as:

\begin{eqnarray} & &
\rho_c(\vec r)=-\sum_{\vec g,\mu \in c,\nu}P_{\nu\vec g\mu\vec 0}
\phi_{\mu}(\vec r-\vec A_{\mu})
\phi_{\nu}(\vec r - \vec A_{\nu}-\vec g)
\end{eqnarray}

The expression which needs to be differentiated has the structure

\begin{eqnarray}
I_{l\mu\vec 0\nu\vec g}^m
=\int \phi_{\mu}(\vec r-\vec A_{\mu})
\phi_{\nu}(\vec r - \vec A_{\nu}-\vec g)
X_l^m(\vec r-\vec A_{\mu}){\rm d^3r}
\end{eqnarray}

We can thus also write

\begin{eqnarray}
\eta_l^m(\rho_c;\vec A_{c})=
-\sum_{\vec g,\mu \in c,\nu}P_{\nu\vec g\mu\vec 0}
I_{l\mu\vec 0\nu\vec g}^m
\end{eqnarray}

When computing the derivative with respect to the cell parameter,
this derivative is required for the expression $I_{l\mu\vec 0\nu\vec g}^m$.
The only dependence on the cell parameter is via the nuclear 
coordinates $\vec A_{\mu}$ and $\vec A_{\nu}+ \vec g$. Thus,
the derivative with respect to the cell parameters $a_{ij}$ is
obtained as 

\begin{eqnarray}
\frac{\partial I_{l\mu\vec 0\nu\vec g}^m}{\partial a_{ij}}=
\frac{\partial I_{l\mu\vec 0\nu\vec g}^m}{\partial A_{\mu,j}}
\left( f_{\mu,i}-f_{\nu,i}-n_{\vec g,i} \right)
\end{eqnarray}

\subsection{Field integrals}

They are defined as

\begin{eqnarray} & &
M_{l\mu\vec 0\nu\vec gc}^m= Z_l^m(\hat{\vec A_{c}})\int
\phi_{\mu}(\vec r-\vec A_{\mu})
\phi_{\nu}(\vec r - \vec A_{\nu}-\vec g)
\bigg[\Phi(\vec r-\vec A_{c})-\sum_{\vec h}^{pen}
\frac{1}{|\vec r-\vec A_{c}-\vec h|}\bigg]{\rm d^3r}
\end{eqnarray}

with $Z_l^m(\hat {\vec A_{c}})$ being the spherical gradient operator
(Ref. \onlinecite{VicCoulomb}). 
The penetration depth $pen$ is a certain threshold for which the
integrals are evaluated exactly \cite{VicCoulomb,Manual03}.

Similar to the nuclear attraction
integrals, this integral also requires derivatives with respect
to $\vec A_{\mu}$, $\vec A_{\nu}+\vec g$ and $\vec A_{c}$, $\vec h$
and the derivatives
with respect to $V$, $\gamma$ and $\vec K$ because of the Ewald
potential.
The derivatives with respect to $A_{\mu}$ and $A_{\nu}+\vec g$
are already available, and the derivatives with respect to
$\vec A_{c}$ are obtained from translational invariance.
We obtain

\begin{eqnarray} & &
\frac{\partial M_{l\mu\vec 0\nu\vec gc}^m}{\partial a_{ij}}=
\frac{\partial M_{l\mu\vec 0\nu\vec gc}^m}{\partial A_{\mu,j}}
f_{\mu,i}+
\frac{\partial M_{l\mu\vec 0\nu\vec gc}^m}{\partial A_{\nu,j}}
(f_{\nu,i}+n_{{\vec g},i})-
\left(\frac{\partial M_{l\mu\vec 0\nu\vec gc}^m}{\partial A_{\mu,j}}
+\frac{\partial M_{l\mu\vec 0\nu\vec gc}^m}{\partial A_{\nu,j}}\right)
f_{c,i}
\nonumber \\ & &
+Z_l^m(\hat{\vec A_{c}})\int
\phi_{\mu}(\vec r-\vec A_{\mu})
\phi_{\nu}(\vec r - \vec A_{\nu}-\vec g)
\bigg[\sum_{\vec h\ne \vec 0} 
\frac{\partial \Phi_{\vec h}(\vec r-\vec A_{c})}{\partial {h_j}}
n_{\vec h,i}-\sum_{\vec h}^{pen} \frac{\partial}
{\partial {h_j}}\frac{1}{{|\vec r-\vec A_{c}-\vec h}|}n_{\vec h,i}
\bigg]{\rm d^3r}
\nonumber \\ & &
+\frac{\partial M_{l\mu\vec 0\nu\vec gc}^m}{\partial V}
\frac{\partial V}{a_{ij}}
+\frac{\partial M_{l\mu\vec 0\nu\vec gc}^m}{\partial \gamma}
\frac{\partial \gamma}{\partial a_{ij}}
+\sum_t \frac{\partial M_{l\mu\vec 0\nu\vec gc}^m}{\partial K_t}
\frac{\partial K_t}{\partial a_{ij}}
\end{eqnarray}

Similar to the nuclear attraction integrals, we can exploit translational
invariance:

\begin{eqnarray} & &
\int \phi_{\mu}(\vec r-\vec A_{\mu}) \phi_{\nu}(\vec r - \vec A_{\nu}-\vec g)
\frac{\partial \Phi_{\vec h}(\vec r-\vec A_{c})}{\partial {h_j}}
n_{\vec h,i}{\rm d^3r}= \nonumber \\ & &
-\int \frac{\partial \phi_{\mu}(\vec r-\vec A_{\mu})}{\partial A_{\mu,j}}
\phi_{\nu}(\vec r - \vec A_{\nu}-\vec g)
\Phi_{\vec h}(\vec r-\vec A_{c}) n_{\vec h,i}{\rm d^3r} \nonumber \\ & &
-\int \phi_{\mu}(\vec r-\vec A_{\mu})
\frac{\partial \phi_{\nu}(\vec r - \vec A_{\nu}-\vec g)}{\partial A_{\nu,j}}
\Phi_{\vec h}(\vec r-\vec A_{c}) n_{\vec h,i}{\rm d^3r}
\end{eqnarray}

and

\begin{eqnarray} & &
\int
\phi_{\mu}(\vec r-\vec A_{\mu})
\phi_{\nu}(\vec r - \vec A_{\nu}-\vec g)
\frac{\partial}
{\partial {h_j}}\frac{1}{{|\vec r-\vec A_{c}-\vec h}|}n_{\vec h,i}{\rm d^3r}
= \nonumber \\ & &
-\int
\frac{\partial \phi_{\mu}(\vec r-\vec A_{\mu})}{\partial A_{\mu,j}}
\phi_{\nu}(\vec r - \vec A_{\nu}-\vec g)
\frac{1}{{|\vec r-\vec A_{c}-\vec h}|}n_{\vec h,i}{\rm d^3r}
-\int \phi_{\mu}(\vec r-\vec A_{\mu})
\frac{\partial \phi_{\nu}(\vec r - \vec A_{\nu}-\vec g)}{\partial A_{\nu,j}}
\frac{1}{{|\vec r-\vec A_{c}-\vec h}|}n_{\vec h,i}{\rm d^3r}
\end{eqnarray}

\subsection{Spheropole}

The spheropole is a correction required because 
the charge distribution is approximated by
a model charge distribution in the long range \cite{VicCoulomb}.

\begin{eqnarray} & &
Q=\sum_{c}Q_c=
\sum_{c}\frac{2\pi}{3V}\int(\rho_c(\vec r)-\rho_c^{\rm model}(\vec r))
|\vec r|^2 {\rm d^3r}= \nonumber \\ & &
\frac{2\pi}{3V}\sum_{c}\sum_{\mu \in c \nu \vec g} P_{\nu\vec g\mu\vec 0}\int
\left[-\phi_{\mu}(\vec r-\vec A_{\mu}) 
       \phi_{\nu}(\vec r-\vec A_{\nu}-\vec g) 
+ \right.
\nonumber \\ & & \left.
\int \phi_{\mu}(\vec r \: '-\vec A_{\mu}) 
       \phi_{\nu}(\vec r\:  '-\vec A_{\nu}-\vec g)
X_l^m(\vec r \: '-\vec A_{\mu}) {\rm d^3r \: '} 
\delta_l^m(\vec A_{\mu},\vec r)\right] |\vec r|^2 {\rm d^3r}
\end{eqnarray}

$\delta_{l}^{m}(\vec {A_c},\vec r)$ is a Gaussian representation of
a unit point multipole\cite{VicCoulomb}.

There are thus dependencies on $\vec A_{\mu}$, $\vec A_{\nu}+ \vec g$, 
and the volume $V$. We need to compute

\begin{eqnarray} & &
\sum_{c}\sum_{\mu \in c \nu \vec g} P_{\nu\vec g\mu\vec 0}
\frac{\partial}{\partial a_{ij}}
\frac{2\pi}{3V}\int
\left[
-\phi_{\mu}(\vec r-\vec A_{\mu}) \phi_{\nu}(\vec r-\vec A_{\nu}-\vec g)
+ \right.
\nonumber \\ & & \left.
\int \phi_{\mu}(\vec r \: '-\vec A_{\mu}) \phi_{\nu}(\vec r\:  '-\vec 
A_{\nu}-\vec g)
X_l^m(\vec r \: '-\vec A_{\mu}) {\rm d^3r \: '} 
\delta_l^m(\vec A_{\mu},\vec r)\right] |\vec r|^2 {\rm d^3r}= 
\nonumber \\ & &
\sum_{c}\sum_{\mu \in c \nu \vec g} P_{\nu\vec g\mu\vec 0}\frac{2\pi}{3V}
\frac{\partial}{\partial A_{\mu,j}}
\int
\left[-\phi_{\mu}(\vec r-\vec A_{\mu}) 
\phi_{\nu}(\vec r-\vec A_{\nu}-\vec g)  \right.
 + \nonumber \\ & & \left.
 \int \phi_{\mu}(\vec r \: '-\vec A_{\mu}) \phi_{\nu}(\vec r\:  '-\vec 
A_{\nu}-\vec g)
X_l^m(\vec r \: '-\vec A_{\mu}) {\rm d^3r \: '} 
\delta_l^m(\vec A_{\mu},\vec r)\right] |\vec r|^2 {\rm d^3r} 
\left( f_{\mu,i}-f_{\nu,i}-n_{\vec g,i}\right)-\frac{Q}{V}\frac{\partial V}{\partial a_{ij}}
\end{eqnarray}

\subsection{Bielectronic integrals}

These are integrals of the type

\begin{eqnarray} & &
B_{\mu\vec 0\nu\vec g\tau\vec n \sigma\vec n+\vec h}= \nonumber \\ & &
\!\int\!\frac {\phi_{\mu}(\vec r\!-\!\vec A_{\mu})
\phi_{\nu}(\vec r\!-\!\vec A_{\nu}-\vec {g})
\phi_{\tau}(\!\vec r\:'\!-\!\!\vec A_{\tau}-\vec n)
\phi_{\sigma}(\vec r\:'\!\!-\!\!\vec A_{\sigma}-\vec n-\vec h)}
{|\vec r-\vec r \: '|} {\rm d^3r \: d^3r'}
\end{eqnarray}

The derivative with respect to the cell parameters $a_{ij}$ is
straightforward, as the only dependence on the cell parameter is
via the position of the centers and thus the nuclear gradients only
need to be multiplied with a factor:

\begin{eqnarray} 
\frac{\partial B_{\mu\vec 0\nu\vec g\tau\vec n \sigma\vec n+\vec h}}
{\partial a_{ij}}=
\frac{\partial B_{\mu\vec 0\nu\vec g\tau\vec n \sigma\vec n+\vec h}}
{\partial A_{\mu,j}}f_{\mu,i}+
\frac{\partial B_{\mu\vec 0\nu\vec g\tau\vec n \sigma\vec n+\vec h}}
{\partial A_{\nu,j}}(f_{\nu,i}+n_{\vec g,i})+ \\
\frac{\partial B_{\mu\vec 0\nu\vec g\tau\vec n \sigma\vec n+\vec h}}
{\partial A_{\tau ,j}}(f_{\tau,i}+n_{\vec n,i})+ 
\frac{\partial B_{\mu\vec 0\nu\vec g\tau\vec n \sigma\vec n+\vec h}}
{\partial A_{\sigma,j}}(f_{\sigma,i}+n_{\vec n,i}+n_{\vec h,i})
\end{eqnarray}

As in reference \onlinecite{CPCarticle}, we define a Coulomb integral
as follows:

\begin{eqnarray} & &
C_{\mu\vec 0\nu\vec g\tau\vec 0\sigma \vec h}
= 
\sum_{\vec n}^{pen}B_{\mu\vec 0\nu\vec g\tau\vec n \sigma\vec n+\vec h}
\end{eqnarray}

and an exchange integral:

\begin{eqnarray}
& &
X_{\mu\vec 0\nu\vec g\tau\vec 0\sigma \vec h}
= 
\sum_{\vec n}
B_{\mu\vec 0\tau\vec n \nu\vec g\sigma\vec n+\vec h} 
\end{eqnarray}

Using this notation, one summation over the lattice vectors
is already included.

\subsection{Energy weighted density matrix}

The contributions of the energy weighted density matrix to the
derivative with respect
to $a_{ij}$: 

\begin{eqnarray} & &
\frac{\partial S_{\mu\vec 0\nu\vec g}}{\partial \vec A_i}  \int_{BZ} 
 \exp({\rm i}\vec k\vec g)\sum_j 
\left\{ \right.
a^{\uparrow}_{\nu j}(\vec k)a^{*\uparrow}_{\mu j}(\vec k)
(\epsilon^{\uparrow}_j(\vec k)+Q)
\Theta(\epsilon_F-\epsilon^{\uparrow}_j(\vec k)-Q) \nonumber \\ & & +
a^{\downarrow}_{\nu j}(\vec k)a^{*\downarrow}_{\mu j}(\vec k)
(\epsilon^{\downarrow}_j(\vec k)+Q)
\Theta(\epsilon_F-\epsilon^{\downarrow}_j(\vec k)-Q) \left. \right\}
{\rm d^3k}
\nonumber \\ & & 
=\frac{\partial S_{\mu\vec 0\nu\vec g}}{\partial \vec A_i}  \int_{BZ} \cdots
\end{eqnarray}

require a different prefactor, compared to the case of the gradients
with respect to nuclear positions:

\begin{eqnarray} & &
\frac{\partial S_{\mu\vec 0\nu\vec g}}{\partial a_{ij}}\int_{BZ} \cdots=
\frac{\partial S_{\mu\vec 0\nu\vec g}}{\partial A_{\mu,j}}
\left( f_{\mu,i}-f_{\nu,i}-n_{\vec g,i}\right) 
\int_{BZ} \cdots
\end{eqnarray}

\section{Total energy and gradient}
\label{Calculationofderivatives}

\subsection{Total energy}

The total energy, in the notation of reference \onlinecite{CPCarticle},
is obtained as follows:

\begin{eqnarray} & & 
E^{\rm total}=E^{\rm kinetic}+E^{\rm NN}+E^{\rm coul-nuc}+E^{\rm coul-el}
+E^{\rm exch-el}=\nonumber \\ & & 
=\sum_{a,b} Z_a Z_b \Phi(\vec A_{b}-\vec A_{a})
   +\sum_{\vec g,\mu,\nu}P_{\nu\vec g\mu\vec 0}T_{\mu\vec 0\nu\vec g}
\nonumber \\ & & 
-\sum_{\vec g,\mu,\nu} 
P_{\nu\vec g\mu\vec 0}\sum_{a}Z_a\int
\phi_{\mu}(\vec r-\vec A_{\mu})
\phi_{\nu}(\vec r-\vec A_{\nu}-\vec g)
\Phi(\vec r-\vec A_{a}){\rm d^3r}
\nonumber \\ & & 
+\frac{1}{2}\sum_{\vec g,\mu,\nu} P_{\nu\vec g\mu\vec 0}
\bigg(-QS_{\mu\vec 0\nu\vec g}+\sum_{\vec h,\tau,\sigma}
P_{\sigma\vec h\tau\vec 0}
C_{\mu\vec 0\nu\vec g \tau\vec 0\sigma\vec h}
-\sum_c \sum_{l=0}^{L}\sum_{m=-l}^{l}\eta_l^m(\rho_c;\vec A_c)
M_{l\mu\vec 0\nu\vec gc}^m\bigg)\nonumber \\ & & 
-\frac{1}{2}\sum_{\vec g,\mu,\nu}P^{\uparrow}_{\nu\vec g\mu\vec 0}
\sum_{\vec h,\tau,\sigma}P^{\uparrow}_{\sigma\vec h\tau\vec 0}
X_{\mu\vec 0\nu\vec g\tau\vec 0\sigma\vec h}
-\frac{1}{2}\sum_{\vec g,\mu,\nu}P^{\downarrow}_{\nu\vec g\mu\vec 0}
\sum_{\vec h,\tau,\sigma}P^{\downarrow}_{\sigma\vec h\tau\vec 0}
X_{\mu\vec 0\nu\vec g\tau\vec 0\sigma\vec h}
\end{eqnarray}

\subsection{Gradient of the total energy}

\label{Gradienttotenysection}

The gradient with respect to the cell parameter is a combination of
the nuclear gradients, with appropriate factors, and new derivatives
with respect to parameters such as $V$, $\gamma$ and $\vec K$ and
their derivatives with respect to the cell parameters.

\begin{eqnarray}
\label{Forcecellparametergleichung} & &
F_{a_{ij}}=-\frac{\partial E^{\rm total}}{\partial a_{ij}}=\nonumber \\ & &
-\frac{\partial E^{\rm NN}}{\partial a_{ij}}
-\sum_{\vec g,\mu,\nu}P_{\nu\vec g\mu\vec 0}\frac{\partial 
T_{\mu\vec 0\nu\vec g}}
{\partial a_{ij}}
-\sum_{\vec g,\mu,\nu} P_{\nu\vec g\mu\vec 0}\frac{\partial 
N_{\mu\vec0\nu\vec g}}{a_{ij}}
\nonumber \\ & &
-\frac{1}{2}\sum_{\vec g,\mu,\nu} P_{\nu\vec g\mu\vec 0}
\bigg\{-S_{\mu\vec 0\nu\vec g}
\sum_{c}
\sum_{\vec h,\sigma,\tau \in c}P_{\sigma \vec h \tau \vec 0}
\frac{\partial}{\partial a_{ij}} \frac{2\pi}{3V} \int\bigg[
-\phi_{\tau}(\vec r-\vec A_{\tau})
\phi_{\sigma}(\vec r-\vec A_{\sigma}-\vec {h})
\nonumber \\ & &
+\sum_{l=0}^L \sum_{m=-l}^l\int 
\phi_{\tau}(\vec r \: '-\vec A_{\tau})
\phi_{\sigma}(\vec r \: '-\vec A_{\sigma}-\vec h)
X_l^m(\vec r \: '-\vec A_c){\rm d^3r'}
\delta_l^m(\vec A_c,\vec r)\bigg]r^2 {\rm d^3r} \nonumber\\ & &
+\sum_{\tau,\sigma}P_{\sigma\vec h\tau\vec 0}
\frac{\partial C_{\mu\vec 0\nu\vec g \tau\vec 0\sigma\vec h}}
{\partial a_{ij}}
-\sum_c \sum_{l=0}^{L}\sum_{m=-l}^{l}\sum_{\vec h,\tau \in  c, \sigma}
P_{\sigma\vec h \tau \vec 0} \nonumber \\ & & 
\frac{\partial}{\partial a_{ij}}\bigg[\int
\phi_{\tau}(\vec r-\vec A_{\tau})
\phi_{\sigma}(\vec r - \vec A_{\sigma}-\vec h)
X_l^m(\vec r-\vec {A_c}){\rm d^3r} \
 M_{l\mu\vec 0\nu\vec g c}^m\bigg]\bigg\}
\nonumber \\ & &
+\frac{1}{2}\sum_{\vec g,\mu,\nu} P^{\uparrow}_{\nu\vec g\mu\vec 0}
\sum_{\vec h,\tau,\sigma}P^{\uparrow}_{\sigma\vec h\tau\vec 0}
\frac{\partial X_{\mu\vec 0\nu\vec g\tau\vec 0\sigma\vec h}}
{\partial a_{ij}} 
+\frac{1}{2}\sum_{\vec g,\mu,\nu} P^{\downarrow}_{\nu\vec g\mu\vec 0}
\sum_{\vec h,\tau,\sigma}P^{\downarrow}_{\sigma\vec h\tau\vec 0}
\frac{\partial X_{\mu\vec 0\nu\vec g\tau\vec 0\sigma\vec h}}
{\partial a_{ij}} 
\nonumber \\ & &
-\sum_{\vec g,\mu,\nu}
\frac{\partial S_{\mu\vec 0\nu\vec g}}{\partial a_{ij}}  \int_{BZ} 
 \exp({\rm i}\vec k\vec g)\sum_n
\left\{ \right.
a^{\uparrow}_{\nu n}(\vec k)a^{*\uparrow}_{\mu n}(\vec k)
(\epsilon^{\uparrow}_n(\vec k)+Q)
\Theta(\epsilon_F-\epsilon^{\uparrow}_n(\vec k)-Q) \nonumber \\ & & +
a^{\downarrow}_{\nu n}(\vec k)a^{*\downarrow}_{\mu n}(\vec k)
(\epsilon^{\downarrow}_n(\vec k)+Q)
\Theta(\epsilon_F-\epsilon^{\downarrow}_n(\vec k)-Q) \left. \right\}
{\rm d^3k}
\end{eqnarray}

\section{Examples}

In this section, we give some numerical examples of the accuracy of
the gradients. First, we consider MgO, at a lattice constant close
to the equilibrium lattice constant. In table \ref{MgOITOL}, numerical and
analytical gradients are compared, for various values of the "ITOL" parameters
controlling the accuracy of the calculation of the integrals. As was
explained in reference \onlinecite{IJQC}, certain parameters (ITOL2, ITOL4
and ITOL5) can introduce an asymmetry in the evaluation of the integrals,
which results in inaccuracies in the gradients. Exactly the same 
holds for
gradients with respect to the cell parameter. The accuracy for the
default parameters is about 2$\times 10^{-4}$ a.u. which should be good enough
for practical purposes,  and by increasing the
values of these parameters, the error is reduced to $10^{-5}$ a.u. 

In table \ref{MgOsupercells}, 
various MgO supercells from $1\times 1\times 1$ (i.e. 
with one magnesium and one oxygen atom in the primitive cell) up to
$5\times 5 \times 5$ have been considered (i.e. with 125 magnesium and
125 oxygen atoms in the primitive cell). The results demonstrate the
high stability of the gradient when larger cells are used.

In table \ref{numanaexamples}, further examples illustrate the accuracy
of the gradients. For various systems, including magnetic ones, the 
analytical and the numerical gradients agree reasonably well. It is again
demonstrated that increasing the ITOL-parameters leads to more accurate
gradients. Also, the stability with respect to the supercell size
is illustrated, for Al$_2$O$_3$.

In table \ref{numanageometry}, the total energy and the 
analytical gradient are displayed, around the equilibrium structure. 
We note that in all cases, the gradient changes sign around the equilibrium:
for example, for MgO, the energy has its minimum between 4.18 and 4.19 \AA,
and also the analytical gradient changes its sign. Similar, in the
other systems considered, the minimum of the energy and the zero of the
analytical gradient agree to 0.01 \AA, at least. 

This is also
demonstrated in table \ref{numanaoptimum}, where the geometry has
been optimized according to the minimum in energy, or the vanishing
of the cell gradient (i.e. the minimum of the energy was determined, up
to an accuracy of 0.001 \AA, and similarly, the geometry with
the smallest value for the gradients was determined, up to an accuracy
of 0.001 \AA). It turns out that the two minima differ at most by 0.004 \AA,
which is probably lower than the noise by the other parameters (basis set,
choice of FIXINDEX parameter\cite{Manual03} and so on).
Note that these calculations were done with the fractional
coordinates of the atoms held fixed, by simply varying the cell
parameters (an automatic optimizer which 
optimizes the cell dimensions and the nuclear positions simultaneously,
using analytical gradients, is not yet implemented in the CRYSTAL code).

Finally, in table \ref{CPUtable}, the CPU times are displayed. The
calculations were performed on a single CPU of a Compaq ES45, with a
clock rate of 1 GHz. It is probably best to compare the CPU time
for the integrals with the time for the gradients, as the code is 
somewhat similar for these two tasks.
At present, the CPU time for all the gradients
(nuclear and cell gradients) is roughly ten times the CPU time
for the integrals. This ratio is expected to be the upper limit as
the gradient code is not yet fully optimized. However, the
calculation of numerical gradients scales with the number of parameters
to be optimized, because at least one more energy point is necessary
for one additional numerical derivative. Thus, if there are enough 
geometrical parameters, the analytical gradients should be clearly favorable.

For the MgO supercells, one can also analyze the CPU times for
the integrals and the self-consistent field procedure 
as a function of the system size. When dividing by the number of
iterations (which is 14, 14, 15, 15 and 18 for the cells from
size $1 \times 1 \times 1$ up to $5 \times 5 \times 5$), 
the CPU time per iteration scales roughly with the third
power of the system size which is to be expected as the diagonalization
scales with this power. The integrals scale with a somewhat lower 
exponent (less than two), due to the fact that more and more
of the bielectronic integrals of the larger cells are not 
evaluated exactly, but with the help of a multipolar expansion.

\section{Conclusion}

A formalism for the calculation of the analytical gradient 
of the Hartree-Fock energy, with respect
to the cell parameter, has been presented and implemented in the code
CRYSTAL, for the case of systems periodic
in three dimensions. The implementation
includes the cases of spin-restricted and unrestricted polarization.
It has been 
shown that a high accuracy can be achieved. Future developments such
as a full structural
optimization with the help of analytical gradients now become feasible.

\section{Acknowledgement}
It is a great pleasure for
K. D. to  thank the Theoretical Chemistry Group (Torino) for
the hospitality during the time spent in Torino. The calculations
were performed on a Compaq ES45
(computer center of the TU Braunschweig),
and on Linux PCs (Torino).

\newpage
\onecolumn

\begin{table}
\begin{center}
\caption{Fcc MgO, at a lattice constant of 4.25 \AA. The accuracy of
the analytical gradient as a function of the truncation parameters
("ITOL"-parameters) is displayed. Basis sets of the size $[3s2p]$ 
were used for Mg and O.}
\label{MgOITOL}
\begin{tabular}{ccc} 
 ITOL & analytical derivative  & numerical derivative \\
 & $[E_h/a_0]$ & $[E_h/a_0]$\\
6 6 6 6 12 (default)     & -0.012737 & -0.012555 \\
8 8 8 8 14     & -0.012589 & -0.012533 \\
10 10 10 10 16 & -0.012522 & -0.012471 \\
10 10 10 16 16 & -0.012496 & -0.012482 \\
10 12 10 16 16 & -0.012505 & -0.012503 \\
\end{tabular}
\end{center}
\end{table}

\begin{table}
\begin{center}
\caption{Fcc MgO, as in table \ref{MgOITOL}. The analytical gradient is
computed as a function of the supercell size to demonstrate the numerical
stability (from 2 to 250 atoms per cell). 
The default ITOL-parameters (6 6 6 6 12) are used.}
\label{MgOsupercells}
\begin{tabular}{ccc} 
supercell size & total energy/MgO unit & analytical derivative \\
 & $[E_h]$ & $[E_h/a_0$]\\
1$\times$1$\times$1 & -274.6635207   & -0.01273658 \\
2$\times$2$\times$2 & -274.6635204   & -0.01273664 \\
3$\times$3$\times$3 & -274.6635205   & -0.01273668 \\
4$\times$4$\times$4 & -274.6635204   & -0.01273665 \\
5$\times$5$\times$5 & -274.6635204   & -0.01273665 \\
\end{tabular}
\end{center}
\end{table}

\begin{table}
\begin{center}
\caption{Other examples for a comparison of analytical and numerical
gradient, including ferromagnetic (FM) and antiferromagnetic (AF) states.
If not stated otherwise, the default ITOL parameters
are used. The basis sets are in the range from $[2s1p]$ for H in urea, up
to $[5s4p2d]$ for the transition metals.}
\label{numanaexamples}
\begin{tabular}{cccccc} 
system & space & cell parameters & component & analytical derivative & numerical derivative\\
& group & [\AA] & & $[E_h/a_0]$ & $[E_h/a_0]$\\
Al$_2$O$_3$ & 
167 & 4.7602, 12.9933 &
$-\frac{\partial E}{a_{1x}}$ & -0.19630 
(2$\times$2$\times$2 cell: -.19630) & -0.19625 \\ \\
Al$_2$O$_3$ & 167 & 4.7602, 12.9933 &
$-\frac{\partial E}{a_{1z}}$ & -0.06366 
(2$\times$2$\times$2 cell: -.06366) & -0.06361 \\ \\
Urea & 113 & 5.565, 4.684 &
$-\frac{\partial E}{a_{1x}}$        & -0.01501 & -0.01475 \\ \\
Urea & 113 & 5.565, 4.684  &
$-\frac{\partial E}{a_{3z}}$        & -0.02495 & -0.02516 \\ \\
NiO, FM & 225 & 4.20 & $-\frac{\partial E}{a_{1z}}$ & 0.00595 & 0.00656 \\ \\
NiO, FM (ITOL: 10 12 10 16 16) & 225 & 4.20 & $-\frac{\partial E}{a_{1z}}$ &
0.00591 & 0.00592 \\ \\
NiO, AF & 225 & 4.20 & $-\frac{\partial E}{a_{3z}}$
& 0.01111 & 0.01234\\ \\
NiO, AF (ITOL: 10 12 10 16 16) & 225 & 4.20  & $-\frac{\partial E}{a_{3z}}$
& 0.01094 & 0.01109\\ \\
KMnF$_3$, FM & 221 & 4.19  & $-\frac{\partial E}{a_{1x}}$ &
0.01043 & 0.01095 \\
\end{tabular}
\end{center}
\end{table}

\begin{table}
\begin{center}
\caption{Other examples for a comparison of analytical and numerical
gradient. The default ITOL parameters are used.}
\label{numanageometry}
\begin{tabular}{ccccc} 
system & cell parameter & energy & components & gradient \\
& [\AA] & [$E_h$] & $[E_h/a_0]$\\
MgO &   4.18  & -274.664192 & $\frac{\partial}{\partial a_{3x}}$ 
&  8.495$\times 10^{-4}$ \\ 
    &   4.19  & -274.664222 &&  8.103$\times 10^{-5}$ \\
    &   4.20  & -274.664209 && -6.735$\times 10^{-4}$ \\
Urea & 5.52, 4.64 &  -447.683214 & $\frac{\partial}{\partial a_{1x}}$,
 $\frac{\partial}{\partial a_{3z}}$  &
7.057$\times 10^{-4}$, 1.4379$\times 10^{-3}$  \\
     & 5.53, 4.63 &  -447.683158 & & -4.045$\times 10^{-4}$, 6.1033$\times 10^{-3}$ \\
     & 5.53, 4.64 &  -447.683218 & & -7.904$\times 10^{-4}$, 6.649$\times 10^{-4}$   \\
     & 5.53, 4.65 &  -447.683176 & & -1.1725$\times 10^{-3}$, -4.7011$\times 10^{-3}$  \\
     & 5.54, 4.64 &  -447.683166 & & -2.2707$\times 10^{-3}$,  -1.010$\times 10^{-4}$ \\
KMnF$_3$  &  4.28 &  -2047.643166 & $\frac{\partial}{\partial a_{1x}}$ &
                                       4.098$\times 10^{-4}$ \\
          &  4.29 &  -2047.643181 & & -5.754$\times 10^{-4}$ \\
          &  4.30 &  -2047.643141 & & -1.5369$\times 10^{-3}$ \\
\end{tabular}
\end{center}
\end{table}

\begin{table}
\begin{center}
\caption{Optimized structures, using energies or analytical gradients. 
The default ITOL parameters are used. For each compound, 
the upper line refers to the
structure with the lowest energy, and the lower line
to the structure with (practically)
vanishing force. The components of the forces are as in table 
\ref{numanaexamples}.}
\label{numanaoptimum}
\begin{tabular}{ccccc} 
system & geometry & energy & force \\
& [\AA] & [$E_h$] & $[E_h/a_0]$\\
KMnF$_3$ & 4.288 & -2047.643182 & -3.8$\cdot 10^{-4}$ \\
         & 4.284 & -2047.643179 &  1.3$\cdot 10^{-5}$ \\
Urea & 5.525 ; 4.642 & -447.683224 & -1.2$\cdot 10^{-4}$ ; -2.9$\cdot 10^{-5}$\\
     & 5.524 ; 4.642 & -447.683224 & 2.8$\cdot 10^{-5}$ ; 4.8$\cdot 10^{-5}$ \\
Al$_2$O$_3$ & 4.497 ; 12.111 & -1401.048515 & -4.4$\cdot 10^{-4}$ ; -1.3$\cdot 10^{-4}$ \\
 & 4.496; 12.111 & -1401.048515 & 2.9$\cdot 10^{-4}$ ; -2.2$\cdot 10^{-5}$
\end{tabular}
\end{center}
\end{table}

\begin{table}
\begin{center}
\caption{CPU times for the various calculations. The calculations were
performed on a Compaq ES45, 
using a single CPU (1 GHz). The CPU times refer to the part for the integrals
(all the integrals were written to disk), the self-consistent field (SCF)
procedure, and to the calculation of all the gradients (i.e. nuclear 
gradients and cell gradients).}
\label{CPUtable}
\begin{tabular}{cccccc} 
system & number of  & \multicolumn{3}{c}{CPU time, in seconds} \\
 & symmetry operators & integrals & SCF & gradients \\
MgO (1$\times$1$\times$1) & 48 & 2 & 0.5 & 26 \\
MgO (2$\times$2$\times$2) & 48 & 11 & 18 & 152 \\
MgO (3$\times$3$\times$3) & 48 & 55 & 500 & 533 \\
MgO (4$\times$4$\times$4) & 48 & 209 & 6330 & 1662 \\
MgO (5$\times$5$\times$5) & 48 & 670 & 57851 & 4443 \\
Al$_2$O$_3$ (1$\times$1$\times$1) & 12 &  15 & 10 & 184 \\
Al$_2$O$_3$ (2$\times$2$\times$2) & 6 & 544 & 4681 & 3877 \\
Urea & 8 & 29 & 103 & 257 \\
NiO, FM & 48 & 12 & 6 & 128 \\
NiO, AF & 12 & 32 &  220 & 346 \\
KMnF$_3$ & 48 & 27 & 20 & 281 \\ 
\end{tabular}
\end{center}
\end{table}

\end{document}